\documentclass[12pt,preprint]{aastex}

\slugcomment{In press on the Astronomical Journal, February 2003 }

\begin{document}

\title{Narrow band imaging in [OIII] and H$\alpha$ to search for ICPNe in the 
Virgo cluster\footnote{Based on observations collected at Subaru
telescope, which is operated by the National Astronomical Observatory
of Japan, at the UT4 of the VLT, which is operated by the European
Southern Observatory, and at the TNG, which is operated by I.N.A.F.
}}

\author{ 
M. Arnaboldi\altaffilmark{1,2},
K.C. Freeman\altaffilmark{3},
S. Okamura\altaffilmark{4,5},
N. Yasuda\altaffilmark{6},
O. Gerhard\altaffilmark{7},
N.R. Napolitano\altaffilmark{1}, 
M. Pannella\altaffilmark{1}
}
\and
\author{
H. Ando\altaffilmark{6},
M. Doi\altaffilmark{10,5},
H. Furusawa\altaffilmark{8},
M. Hamabe\altaffilmark{11},
M. Kimura\altaffilmark{9,12},
T. Kajino\altaffilmark{6},
Y. Komiyama\altaffilmark{8},
S. Miyazaki\altaffilmark{6},
F. Nakata\altaffilmark{4},
M. Ouchi\altaffilmark{4},
M. Sekiguchi\altaffilmark{9},
K. Shimasaku\altaffilmark{4,5},
M. Yagi\altaffilmark{6}
}

\altaffiltext{1}{I.N.A.F., Osservatorio Astronomico di Capodimonte,
80131 Naples, Italy }
\altaffiltext{2}{I.N.A.F., Osservatorio Astronomico di Pino Torinese,
10025 Pino Torinese, Italy}
\altaffiltext{3}{R.S.A.A., Mt. Stromlo Observatory, 2611 ACT, Australia}
\altaffiltext{4}{Department of Astronomy, University of Tokyo, Bunkyo, Tokyo
113-0033, Japan}
\altaffiltext{5}{Research Center for the Early Universe, University
of Tokyo, Bunkyo, Tokyo 113-0033, Japan}
\altaffiltext{6}{National Astronomical Observatory of Japan, Mitaka, Tokyo
181-8588, Japan}
\altaffiltext{7}{Astronomisches Institut der Universit\"at, CH-4102 Binningen, Switzerland}
\altaffiltext{8}{National Astronomical Observatory of Japan, Hilo, Hawaii
96720. U.S.A.}
\altaffiltext{9}{Institute for Cosmic Ray Research, University of Tokyo,
Kashiwa, Chiba 277-8582, Japan}
\altaffiltext{10}{Institute of Astronomy, University of Tokyo, Mitaka, Tokyo
181-8588, Japan}
\altaffiltext{11}{Department of Mathematical and Physical Sciences,
Japan Women's University, Bunkyo, Tokyo 112-8681, Japan}
\altaffiltext{12}{present address: Department of Astronomy, Kyoto
University, Sakyo-ku, Kyoto 606-8502, Japan}

\begin{abstract}
We have identified intracluster planetary nebulae in a Virgo cluster
core field by imaging with the Subaru Suprime-Cam through two narrow
band filters centered at the redshifted wavelengths of the [OIII]
$\lambda=5007$ \AA\ and the H$\alpha$ $\lambda = 6563$ \AA\ lines;
broad-band images in V and R were acquired to check for 
emission in the adjacent continuum.  Emission line objects in Virgo are
then selected from the two-color diagram ([OIII] -- H$\alpha$) vs.
([OIII] -- (V+R)), which is calibrated using PNe in M84 (Jacoby et al.
1990).  Using both [OIII] and H$\alpha$ allows us to distinguish
bona-fide planetary nebulae from high redshift emission-line galaxies
at the bright end of the [OIII] luminosity function.
Spectroscopic observations of a subsample of these objects were made at
the TNG and at the VLT, in a region around M84 and in an intracluster field
respectively. The observations confirm the efficiency of the combined
[OIII]+H$\alpha$ imaging to identify true PNe.  We also obtained the
first spectrum of an intracluster PN which shows the [OIII] doublet
with S/N$ > 10$ and its H$\alpha$ emission.  From the results
based on the spectroscopic follow-up, we derive a lower limit to the
fraction of the Virgo cluster light contributed by the intracluster
stars at the surveyed position in the cluster core: it amounts to 10\%.
\end{abstract}


\section{Introduction}
The diffuse stellar light in galaxy clusters is now a well established
component of the intracluster (IC) medium (Zwicky 1951; 
see also review by Feldmeier 2002) and the study of its 
properties is a potentially powerful method for probing the formation 
and evolution of cluster of galaxies (Dressler 1984; Miller 1983; Merrit
1984;  Moore et al. 1996;  Moore, Quills \& Bower 2000).

Studies based on direct measurements of the IC surface brightness 
are difficult because the typical surface brightness of the IC light 
is less than 1\% of the sky brightness,
and it is hard to disentangle the light associated with
the halo of the cD from the light associated with the cluster 
(Bernstein et al. 1995).
A new approach came with the discovery of intracluster planetary
nebulae (ICPNe) candidates in the Virgo and Fornax clusters
based on spectroscopy (Arnaboldi et al. 1996)
and imaging in the light of the $\lambda5007$\AA\ line of [OIII] 
(Theuns \& Warren 1997; Ciardullo et al. 1998; 
Feldmeier, Ciardullo, \& Jacoby 1998; Feldmeier 2002; Arnaboldi et
al. 2002). 
Although this technique is effective for detecting ICPNe, it also detects 
other objects with emission lines in this narrow band.  The first 
spectroscopic follow-up (Kudritzki et al. 2000; Freeman et al. 2000) 
carried out on a subsample of these objects showed contamination by 
{\it i)} stars misclassified as emission line objects because of
photometric errors, and 
{\it ii)} high redshift line-emitters, like [OII] starbursts at 
$z \simeq 0.347$, and Ly$\alpha$ objects at $z \simeq 3.13$.
An automatic selection procedure implemented by Arnaboldi et
al. (2002) showed that contamination from continuum objects can be
solved via acquisition of an adequate off-band image, while reducing
contamination from high-z line-emitters is a more serious matter.\\
Arnaboldi et al. (2002) gave an estimate of the 
contamination from high-z line-emitters, of up to 25\% within the 
first magnitude from the bright cut-off of the PN luminosity function 
(PNLF) in the IC fields, based on the spectroscopically confirmed 
sample of ICPNe from Freeman et al. (2000).
Narrow band imaging survey of emission line objects carried out 
on empty fields (Ciardullo et al. 2002a) provides independent
estimates of the fraction of high redshift interlopers. 
Using the same identification techniques as in the Feldmeier et al. (1998) 
ICPN sample, Ciardullo and collaborators determined 
that emission-line emitters at high redshift contribute about 
20\% of the ICPN selected sample in the Virgo cluster. 
Because of their large equivalent width ($W_\lambda > 82$\AA), 
the high redshift interlopers are most probably Ly$\alpha$ emitters 
at redshift 3.13.

In principle, the high redshift line emitters 
can be statistically subtracted from a larger sample of ICPNe candidates in 
Virgo/Fornax, when one wishes to determine the luminous density 
of the diffuse IC light. In this approach one needs to assume that 
the Ly$\alpha$ emitters are not highly clustered.   
This assumption has been questioned by Steidel et al. (2000) and Porciani \& 
Giavalisco (2002): the former have claimed that the L$\alpha$ emitters at 
$z \simeq 3$ are the same population as the Ly-break galaxies at the same 
redshift, while the latter have shown that the Ly-break population 
has a high clustering signal.\\
Recently Ouchi et al. (2002) detected a clear clustering
signal for Ly$\alpha$ emitters at $z=4.86$. This suggests that
Ly$\alpha$ emitters are definitely clustered at $z\simeq3$, since
they are already clustered at $z=4.86$.

To overcome these difficulties, we can use the {\it two} strongest 
emission lines from PNe, i.e. [OIII] $\lambda5007$\AA\ and H$\alpha$, carry
out narrow band imaging with two filters centred on each of the two
lines respectively, require detection in both narrow band images and
absence of continuum emission to identify true Virgo ICPNe.
The H$\alpha$ flux from a PN is typically 3 to 5 times weaker 
than the corresponding flux in the [OIII] line, so we need a large 
telescope with wide field capabilities in order to carry out such an 
imaging survey for the ICPNe in Virgo. 
The Subaru 8.2-meter telescope and the Suprime-Cam mosaic camera
(Miyazaki et al. 2002) represent the best suited instrument to perform 
such a survey of ICPNe in Virgo and trace the cluster 3D shape via 
the bright cut-off of the PN luminosity function.

As discussed by Ciardullo et al. (2002b), imaging in [OIII] and 
H$\alpha$ in late-type galaxies is also efficient in discriminating PN
from HII regions: for candidate objects in the first magnitude of the 
PNLF we can identify compact HII regions, if such objects are 
present, as those with [OIII]$\lambda 5007$ to 
H$\alpha+$[NII] line ratios less than 2. By adopting this
strategy, and by surveying large areas in the Virgo cluster, we may be 
able to find compact intracluster HII region.

Here we present a broad band (V and R bands) and narrow
band imaging surveys of a Virgo IC field at $\alpha = 12:25:47.0$,
$\delta = 12:43:58$ (J2000), which were carried out during the
guaranteed time of
the Suprime Camera commissioning phase, at the Subaru telescope in
March and April 2001, and the subsequent spectroscopic follow-up 
in April 2002 at the TNG 4 meter telescope with DOLORES, 
and at UT4 of the VLT 8 meter telescope with FORS2. 
The ICPN candidates from this imaging survey are presented in 
Okamura et al. (2002).
In Section~2 we describe our observations, data reduction
and the color properties of the selected candidates. In Section~3 we
describe the catalog extraction and discuss the selection
procedure, via simulations of PNe and single line emitters 
in the blank areas of our surveyed field. 
In Section~4 we calibrate our selection criteria on the
published PNe sample of M84 and M86 from Jacoby et al. (1990: JCF90), and 
summarize the properties of our best sample of IC objects in Virgo. In
Section~5 we validate our selection criteria via the spectroscopic
follow-up, and describe an unambiguous spectroscopic confirmation 
of an intracluster PN. 
In Section~6 we discuss our estimate on the IC diffuse light and 
conclusions are presented in Section~7.

\section{Imaging, data reduction and photometric candidates}
In March-April 2001 a field in the Virgo cluster at $\alpha (J2000)=
$12:25:47.0, $\delta(J2000) = $12:43:58, just south of M84-M86, was
observed during the commissioning of the Suprime-Cam 10k$\times$8k
mosaic camera, at the prime focus of the 
Subaru 8.2-meter telescope. The field
of the array covers an area of $34 \times 27$ arcmin of sky, with a pixel
size of $0''.2$.

The field was imaged through two narrow band and two
broad band filters. The two narrow band filters have ($\lambda_c,
\Delta\lambda) = (5021$\AA, $74$\AA) and ($6607$\AA, $101$\AA) {
for the f/1.86 beam at the Subaru prime focus, see Appendix},
corresponding to the [OIII] and H$\alpha$ emissions at the
redshift of the Virgo cluster. The filter widths were chosen to match
the velocity dispersion of galaxies in the Virgo cluster. For the two
broad band filters, we used standard V and R filters. 
The observations log is summarized in Table \ref{log}.
The broad band data were calibrated using Landolt stars. 
The adopted V band calibration for the true extinction 
corrected V magnitude is $m_{Vtrue} = m_{Vinstr} + (28.0 \pm 0.05$). 
No spectrophotometric standard stars were observed during the commissioning 
runs, nonetheless the narrow band images can be calibrated in AB
magnitudes by setting the colors of the continuum objects equal to
zero, as is done in the LALA survey (Rhoads et al. 2000). 
Furthermore, our narrow band [OIII] instrumental
magnitudes can be calibrated in $m(5007)$ magnitudes, as
introduced by Jacoby (1989), by comparison with published 
PN photometry.\\
Removal of instrumental signature, astrometric solution and 
image co-addition were done with a data reduction package developed by the
the Suprime-Cam team.

Once the final images in the four filters are normalized to 
1 sec exposure, the instrumental magnitudes define colors which
depend on the objects' nature. For example, the colors of the
blue stars with a flat spectrum are found to be at 
([OIII] - V) = 4.0, with a similar value for the (H$\alpha$
- R) color; these colors are mostly due to the filter width ratio.  
In the instrumental ([OIII] - H$\alpha$) color,  
blue stars are at ([OIII] - H$\alpha$) = 1; while 
foreground galaxies and red stars are found at ([OIII] - H$\alpha) \geq
1$.   In what follows we will use a combined (V+R)
continuum image, following the procedure adopted by Steidel et al.
(2000).

\section{Catalog extraction and validation of the selection 
procedure\label{color}}
As a first step, we established the limiting magnitudes for the different 
images following the simulation procedure described in Arnaboldi et al. (2002).
We define the {\it limiting magnitude} of our final co-added mosaic image 
as the magnitude at which half of the input sample is retrieved from the 
simulated image, see Arnaboldi et al. (2002) for further details.  
The {\it detection limit} is the magnitude at which the fraction of
retrieved point-like objects becomes zero.
The limiting instrumental magnitudes are:
\\ $m(V+R)_{lim} = -2.3$ , $m([OIII])_{lim} =  1.8$,
$m(H\alpha)_{lim}=  1.2$\\ 
The detection limits are:\\ $m(V+R)_{det} =
-2.0$ , $m([OIII])_{det} =  2.1$, $m(H\alpha)_{det}=  1.5$.\\ 

The images in our dataset, (V+R), [OIII] and H$\alpha$, are 
complete (which corresponds to detections with $S/N \sim 9$) 
down to these instrumental 
magnitudes:\\ $m(V+R)_{com} = -2.8$ , $m([OIII])_{com} =  1.3$,
$m(H\alpha)_{com}=  0.7$\\ 
As we will discuss further in Section 4, the $S/N$ ratio for our V 
and R images is lower than ideal. Even if we use the combined
(V+R) image for our continuum band, our detection and color
selection process is limited more by the $S/N$ ratio of the 
off-band (V+R) image than by the on-band [OIII] image.

At what $m([OIII])$ instrumental magnitude do we expect Virgo PNe to come 
in? Arnaboldi et al. (2002) carried out their ICPN survey 
with a narrow band [OIII] filter width very similar to ours, and
their brightest ICPN candidates are detected at $m_{AB} \sim 23.5$  in
[OIII]. Using the calibration of the V band, and accounting for a
$([OIII] - V) = 4.0$ color, we would expect the brightest ICPNe to come in
at $m([OIII])\sim -0.5$, 1.8 magnitude brighter than our
$m([OIII])_{com}$.

From the ([OIII] -- H$\alpha$) vs. [OIII] color-magnitude diagrams of all the
sources in the field, we would expect PNe to have $([OIII] - H\alpha)
\sim 0.25$, i.e. about 0.75 magnitude bluer than the blue continuum
stars.  This estimate comes from the observed distribution of
the [OIII] to (H$\alpha$+[NII]) line ratios for [OIII]-selected
planetary nebulae (Ciardullo et al. 2002b) in late-type galaxies . 
Because a { luminous} PN's [OIII] to H$\alpha$ ratio varies between 3 and
5, we can expect PNe, and ICPNe, to scatter to bluer colors with
respect to flat blue continuum stars. { We note that excitation and
metallicity may allow ratios as small as 1.}

\begin{figure}
\epsscale{0.8}
\plotone{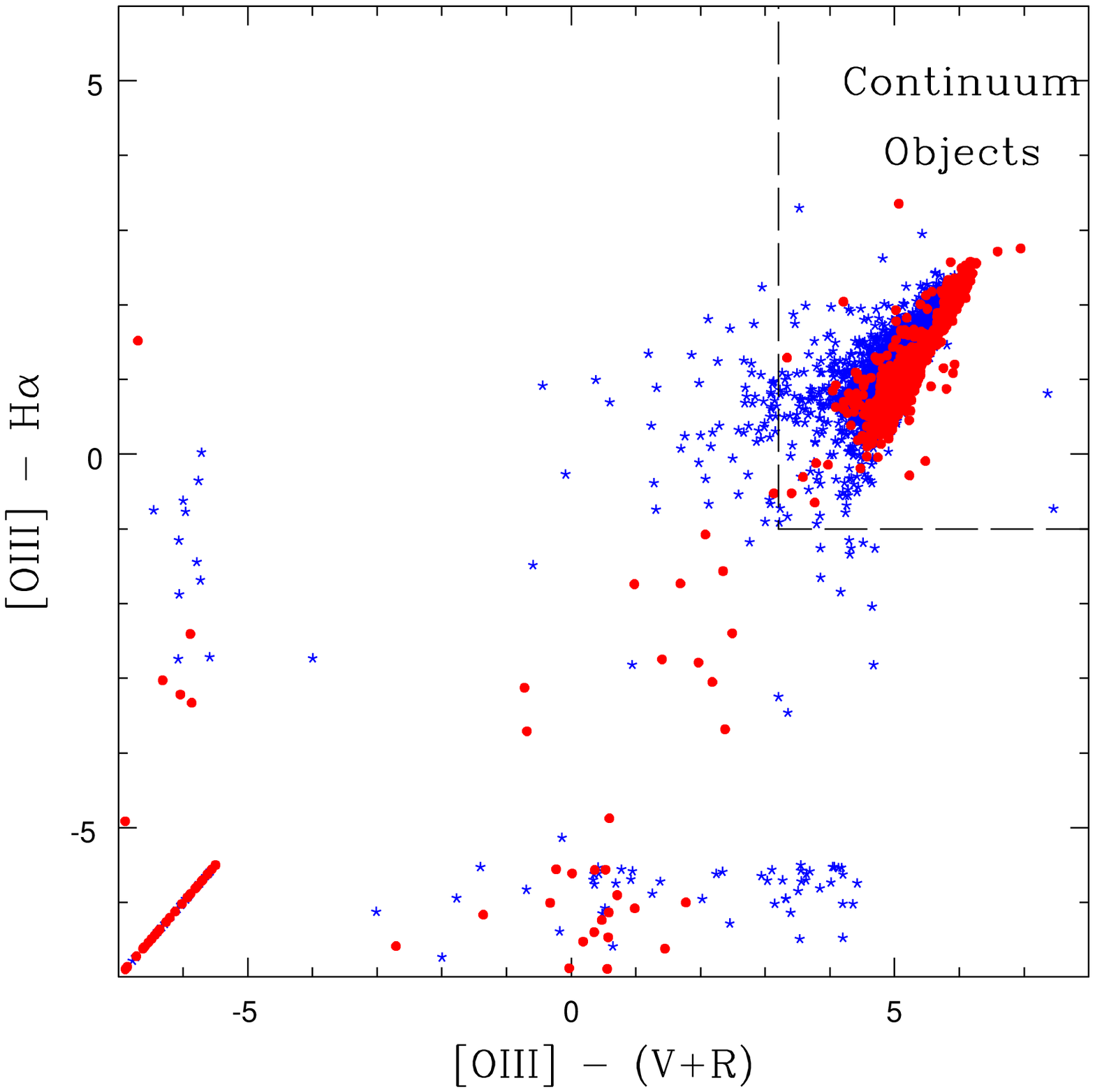}
\caption{Two-color diagram of bright [OIII] detected sources
with $m([OIII]) < -0.5$ in the Subaru field. 
Red dots are point-like objects, blue stars are extended 
objects. Objects aligned with the x-axis are those for which the measured 
H$\alpha$ flux at the [OIII] source position is negative, and their
magnitudes are set arbitrarily to $5$. Objects aligned with the y-axis 
are those for which
the measured (V+R) flux at the [OIII] source position is negative, and
their magnitudes are set arbitrarily to $5$.\label{fig1}}
\end{figure}

Given the limiting magnitude of the [OIII]
image, we see that the long exposure H$\alpha$ image is 
{\it not} deep enough to ensure detection of the H$\alpha$ line 
for all the ICPNe that we may be able to detect in the [OIII] image.
In the first magnitude bin of the observed PNLF, from
$m([OIII])\sim -0.5$ to $m([OIII]) \sim 0.5$,
the depth of the H$\alpha$ image would allow us to detect 
both high and low excitation PN\footnote{A high excitation
PN with $m([OIII]) \sim 0.5$ would have $m(H\alpha) \sim
0.7$, a low excitation PN would have $m(H\alpha) \sim 0.2$. These 
H$\alpha$ magnitudes are both brighter than the completeness 
limit of our H$\alpha$ image}; 
when $m([OIII]) > 0.5$, we loose the high excitation PN.
For an easily detectable ICPN candidate with $m([OIII]) =
1.5$ (which has $S/N > 5$ ), the expected $m$(H$\alpha$) for a
low excitation PN is 1.25, which is now at the limiting magnitude 
of the H$\alpha$ image.\\
From the calibration of the ([OIII] - V) color, the distribution of 
bright sources in the field (see Figure\ref{fig1}), and the simulations of
continuum point-like sources on the (V+R) image, the combined
continuum image is
deep enough for the detection of a continuum object, with a
flat spectrum, and $m([OIII]) = 1.5$.
Spurious detections may arise for those objects with a continuum 
increasing towards the blue, as they may mimic a detection in both narrow 
band images (with lower flux in the H$\alpha$ image), but have no 
detected (V+R) flux because of the higher noise in the continuum sky 
background. We can reduce this problem by selecting only the 
brighter [OIII] sources in our analysis.
We shall discuss this point and the relative $m(5007)$ threshold in
Section~4.

We use SExtractor (Bertin \& Arnouts 1996) to carry out the detection and 
photometry.  Ideally,
the [OIII], H$\alpha$ and (V+R) images would be so deep that we could
make reliable detections or very significant non-detections in these
bands for all potential candidates. We could then detect PNe and
eliminate single-lined high redshift emission-line galaxies and
spurious continuum objects from our sample of candidates, using
independent detections and photometry for the [OIII], H$\alpha$ and
off-band (V+R) images. In reality,
because the [OIII] image is the deepest, the adopted strategy is to use
the [OIII] image to perform the source detection, and then compute
aperture magnitudes for the H$\alpha$ and (V+R) images, at the location of
the [OIII]-detected objects.

{\it Selection procedure --}
As a first step, we defined the criterion to select unresolved (point-like)
objects from the more extended objects. We used the SExtractor 
mag(AUTO) against flux-radius diagram for all the [OIII]-detected sources. 
Point-like sources are found in a locus of points parallel to
the mag(AUTO) axis, with a flux-radius value given by the FWHM of the seeing
disk. From this plot, and the average seeing measured on the [OIII]
image, $\sim 0''.72$, we classified as point-like sources those [OIII]
detection with a flux-radius value less than 3.6 pixels;
sources with larger flux-radius are taken to be extended. 

We use the  two-color diagram  ([OIII] - H$\alpha$) vs. ([OIII] - (V+R))
to select our emission-line objects. The two-color diagram for the bright 
sources ($m([OIII]) < -0.5$) is shown in Figure~\ref{fig1}. 
These bright objects are almost entirely stars and galaxies.
The blue and red stars plus the foreground galaxies (the extended objects) 
populate a well defined area of the two-color plot, centered at 
([OIII] - (V+R)) = 5, and  ([OIII] - H$\alpha$) between -1.0 and 3.0.

We now establish some {\it ad hoc} selection rules for PNe, adopted for
this specific set of images and the detection/photometry process
described above.  Simulated populations of PNe, single-lined emitters
and continuum sources are constructed in the two-color diagram, and are
used to outline the regions inhabited by the different kinds of
objects.  The two bright galaxies M86 (NGC 4406, {
$v_{sys} = - 244$ km s$^{-1}$}) and M84 (NGC 4374,  { $v_{sys} = 1060$ 
km s$^{-1}$}) appear in our Subaru field. These have been
independently surveyed for PNe, by JCF90, and we can evaluate our selection
criteria against the photometry for the JCF90 samples.  The selection
criteria will then be further validated with spectroscopy of the
candidate objects.

{\it Simulated population} -- A population of PNe was represented by a set 
of point-like objects with $m([OIII]) = 1.0$, a  
([OIII]--H$\alpha$) = 0 color\footnote{This corresponds to an [OIII] to 
$H\alpha$+[NII] ratio equal 3. We include [NII] in the evaluation of
the line fluxes because of the H$\alpha$ filter width, which includes
the flux in these lines too.}, and no flux in (V+R).  
These were added to the modeled background image 
for the [OIII] and the H$\alpha$ images (see Arnaboldi et al. 2002 
for additional details).  
Photometry was then performed on these simulated objects, using SExtractor, 
to study their distribution in the two-color plane. Sources were detected 
on the simulated [OIII] image, and aperture photometry was performed at 
the corresponding position on the H$\alpha$ and (V+R) image.

\begin{figure} 
\epsscale{0.8}
\plotone{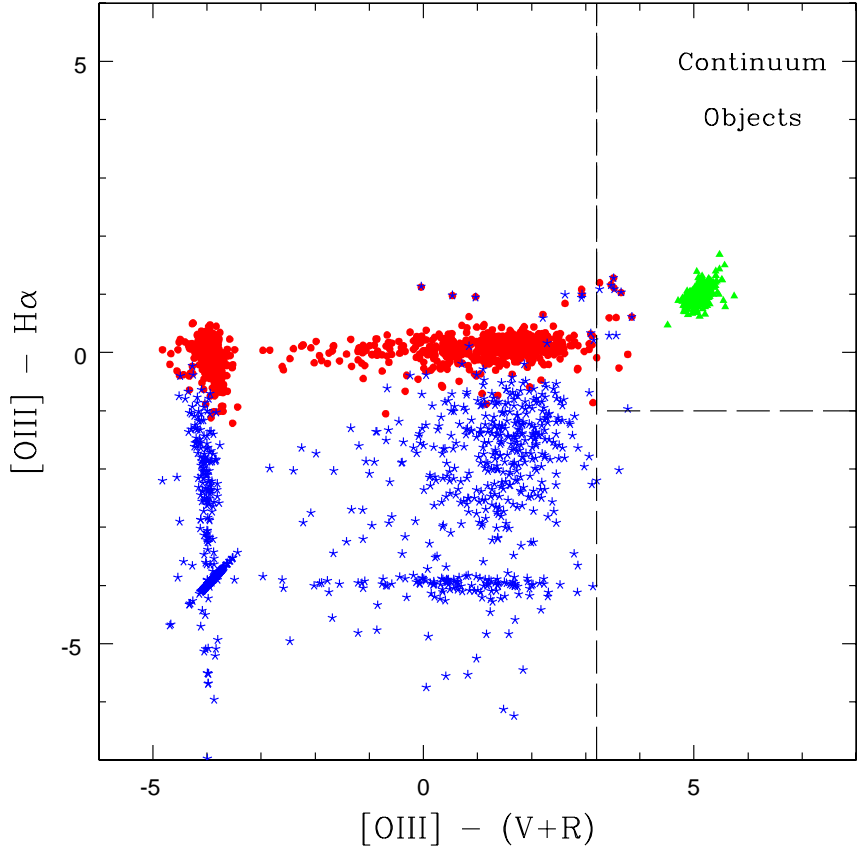}
\caption{Two-color diagram of the simulated population of point-like 
objects. Red dots are PN-like objects with with $([OIII]-H\alpha) =0,\, 
m([OIII])=1.0$, no flux in (V+R).
Blue stars are one-line emitters with $[OIII] = 1.0$, no $H\alpha$, no
(V+R). Green triangles are a point-like population of continuum objects with 
$m([OIII]) = 1.8$, $([OIII] - H\alpha) = 1.0$ and $([OIII] - (V+R)) = 5$. 
Objects which are distributed parallel to the x-axis at 
$([OIII] - H\alpha) \simeq -4.0$ are those for which the measured 
H$\alpha$ flux at the [OIII] source position is negative, and their
magnitudes are set arbitrarily to $5$. 
Objects which are distributed parallel to the y-axis at 
$([OIII] - (V+R)) \simeq -4.0$ are those for which the measured (V+R) 
flux at the 
[OIII] source position is negative, and their magnitudes are set 
arbitrarily to $5$. A negligible fraction of line emitters are scattered
at colors $([OIII] - (V+R)) > 3.2$.
\label{fig2}}
\end{figure}

Figure~\ref{fig2} shows the simulated distribution of a PNe population with 
$m([OIII]) = 1.0$ (i.e. 0.8 mag brighter than the limiting magnitude). 
The measured (V+R) flux is due entirely to noise in the (V+R) image, so
the PNe are distributed in two regions: (i) for some of them,  
the measured ``flux'' in (V+R) is negative, and these objects are arbitrarily 
assigned a value of $m(V+R) = 5$;  (ii) for others, the measured 
(V+R) flux is positive but they lie fainter than detection limit 
$(m(V+R)_{det} = -2)$, { i.e. their $m(V+R) > m(V+R)_{det} $. Therefore
most of our points are situated in the region $(m([OIII]) - m(V+R)) < 1.0 - 
m(V+R)_{det}$, i.e. $(m([OIII]) - m(V+R)) < 3 $. }
Errors in the photometry, halos of bright objects, and CCD defects may cause 
the measured $m(V+R)$ to be brighter than $-2.0$, so a few points may 
scatter into the region $([OIII] - (V+R)) > 3.0$.\\

We then simulated a population of {\it single-lined} emitters, i.e.
with $m([OIII])=1.0$ and no emission in H$\alpha$ and (V+R). A point-like
population is added to the simulated [OIII] background image only, and
aperture photometry is carried out at the position of the detected
[OIII] sources on the H$\alpha$ and (V+R) background. For these objects,
the measured fluxes in H$\alpha$ and in (V+R) are due entirely to noise
in the H$\alpha$ and (V+R) images. When a negative flux is measured,
$m(H\alpha)$ and $m(V+R)$ are set arbitrarily to 5, so objects detected
in [OIII], but with negative fluxes in both H$\alpha$ and (V+R), lie on a
diagonal line in the lower left corner of the two-color diagram.  These
single line emitters lie in a different area from the PNe (see
Figure~\ref{fig2}), although a few may scatter into the region of the 
PNe and so contaminate our sample of candidates.  With a deeper image in
H$\alpha$, matching the limiting magnitude of the [OIII] image, these
objects would not be a significant contaminant, as we would require
detection both in [OIII] and H$\alpha$.

Finally we added to the real image a simulated point-like 
population of continuum emitters with [OIII] = 1.8, and a flat continuum 
spectrum which is given by $([OIII] - H\alpha) = 1$ and $([OIII] - (V+R)) 
= 5$, ( i.e. $m(V+R) = -3.2$), and retrieved their position in the 
two-color diagram.  The simulated objects cluster together at the loci 
where the observed continuum objects lie, see Figure~\ref{fig2}. 

From the two-color diagram of bright sources, and the simulated
populations of PNe and single line emitters, 
we can identify the following selection criteria for objects which are
{\it detected} with SExtractor on the [OIII] image and 
with {\it measured} aperture photometry on the other images, 
at the position of the [OIII] source: 
\begin{enumerate}
\item for stars and continuum extended objects we expect to detect
fluxes in [OIII], H$\alpha$ and (V+R), which translate in following
color selections, based on our simulations: $([OIII]-(V+R)) > 3.2$ and 
$([OIII] - H\alpha ) > -1.0$;
\item for single-lined emitters, either Ly$\alpha$ + Continuum or 
[OII] + Continuum, we expect them to be detected in [OIII] and 
have positive fluxes in (V+R), but to be very weak or have negative 
flux in H$\alpha$, i.e. $m(V+R)\leq -2.8$ and  $([OIII] - H\alpha)
<-1.0$. The $([OIII]-(V+R))$ color would depend on the equivalent width of
the line emission, and therefore be on both side of 
$([OIII]-(V+R)) = 3.2$;
\item for single-lined emitters with no continuum, i.e.
Ly$\alpha$ emitters at $z = 3.13$ and [OII] emitters at $z=0.34$,
we expect them to be bright in [OIII], very weak in both 
H$\alpha$ and (V+R). Objects that appear as
$([OIII] - (V+R)) < 3.2$, $m(V+R)>-2.8$, and $([OIII] - H\alpha) <
-3.0$ are single line emitters, but because of noise some get
scattered up to $([OIII] - H\alpha) < -1.0$;
\item for ICPNe candidates: $([OIII]-(V+R)) < 3.2$,\, $m(V+R) >-2.8$  and 
$-3 < ([OIII] -H\alpha) < 1$\footnote{ Here the range for the 
$([OIII] -H\alpha)$ color accounts for the errors from SExtractor
photometry in [OIII] and H$\alpha$, which cause an intrinsic flux
ratio from 3 to 5 to be scattered to larger values.}. 
Given that for our H$\alpha$ image,  $m(H\alpha)_{com}=  0.7$, 
a high excitation PN with $[OIII] \geq 0.5$ may not have 
a measured H$\alpha$ flux, and therefore it is scattered in the 
region of the two-color diagram where single lined emitters are found.
\end{enumerate}

From the simulations of a single-line-emitting point-like population, some 
contaminants may still be scattered by noise in the regions selected 
for PNe. 
As an additional test, we visually inspect all the selected candidates 
for the presence of an H$\alpha$ source at the position of the detected 
[OIII] source, and only these objects will be considered as PN candidates 
in Virgo.

\section{Comparison with the M86, M84 sample of JCF90}
The bright galaxies M86 and M84, which were surveyed for PNe by JCF90, appear
on our images. The PNe sample of JCF90 were compared with our catalogs of 
detected sources from the deep Subaru images.

\begin{figure}
\epsscale{0.8}
\plotone{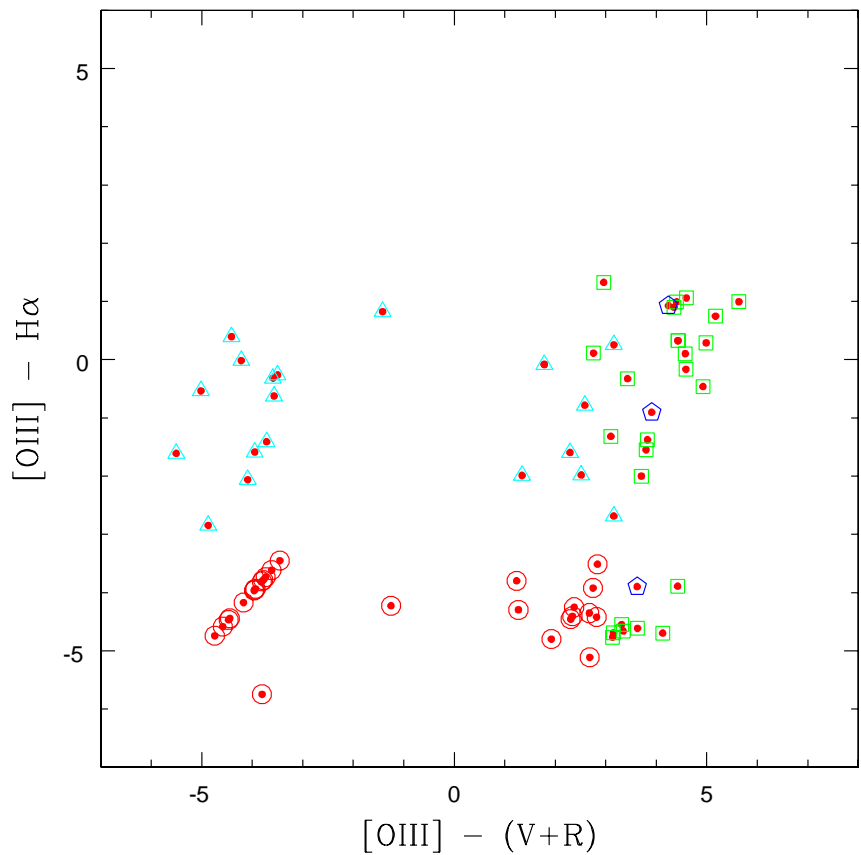}
\caption{Two-color diagram for the M84 matched sample.  Red dots are
[OIII] sources matched to the JCF90 sample. Cyan triangles (19 objects)
indicate the best PNe candidates, red circles (26) indicate the 
single line emitters with $m(V+R) > -2.8$, most likely
high excitation PNe in this situation. Green squares (26) indicates
objects with significant flux in the (V+R) image, i.e. 
$m(V+R) < -2.8$. Blue pentagons (3) indicates objects with unreliable colors.
\label{fig3}}
\end{figure}

For M84, JCF90 reported 102 PNe candidates:  74 were matched with sources 
on the Subaru images, within a separation of $1''.4$. Of the remaining 28
objects, 18 were at  JCF90 magnitudes fainter than 
$m(5007) = 27.7$ (i.e. fainter than our [OIII] completeness limit 
$m(5007) = 27.6$ - see below) and 11 were not detected on the Subaru [OIII]
image.
See Figure~\ref{fig3} for the two-color plot for the matched objects.  
Of the 74 objects, 
\begin{itemize}
\item 19 are detected in [OIII] and have a positive H$\alpha$ flux,
with $m(V+R)>-2.8$ (most likely true PNe), they comply to the criterion
n. 4 ;
\item 26 are detected only in [OIII], have a very faint continuum
($m(V+R)>-2.8$), and have a negative measured 
H$\alpha$ flux; they appear to comply to criterion n. 3.
These sources are most likely high excitation PNe, 
given that we are looking within the light distribution of M84, where 
PNe dominate over the high-z background of line emitters. 
A very small fraction may still be Ly$\alpha$ emitters though,
with no detected continuum and a large equivalent width.
\item 26 objects have $m(V+R)<-2.8$, i.e. significant flux in
the combined (V+R) image. 12 of these fall in the
area of the two-color plot where flat-spectrum continuum objects are found;
7 are single-line emitters according to criterion n. 2, with 
negative flux in H$\alpha$\footnote{ We would expect $ 5 \pm 2$ objects
with detected (V+R), from the Ly$\alpha$ population in the field at 
redshift $z=3.13$ from Steidel et al. (2000).};
6 objects are most likely continuum objects, because of their 
significant flux in (V+R), with a steep continuum towards the blue
as indicated by their measured $([OIII]- H\alpha) < 0.0$ ( while 
continuum objects with flat spectrum have $([OIII]- H\alpha) \simeq 1.0$).
In this set of sources, we unexpectedly find a candidate HII region 
in the M84 field, at $([OIII]- H\alpha) > 1.0\,\, 
\mbox{and}\,\, ([OIII]-(V+R)) \simeq 3$.
Its measured instrumental magnitudes are 
$m([OIII] =-1.03,\, m(H\alpha)= -2.36\,\, \mbox{and}\,\, m(V+R) = -4.0$, all
above the completeness magnitude limits in their bands, and therefore
it is unlikely that this source comes from a peak in the noise.
We may search for compact HII regions in the IC field too,
according to the following selection criteria:
$([OIII] - (V+R)) < 3.2$, $([OIII] - H\alpha) \geq 1.0$, and these candidates
may have some flux in the continuum, i.e. $m(V+R) < -2.8$.
\item The remaining three objects are scattered in the region where 
the continuum objects are found, but their $m(V+R) > -2.8$: these
are the three faintest ($m([OIII])> 1.2$) objects in the matched
sample, and may be their colors are most affected by 
photometric errors. See Figure \ref{fig3} for the two-color plot for the 
M84 matched sources. 
\end{itemize}

\begin{figure}
\epsscale{0.8}
\plotone{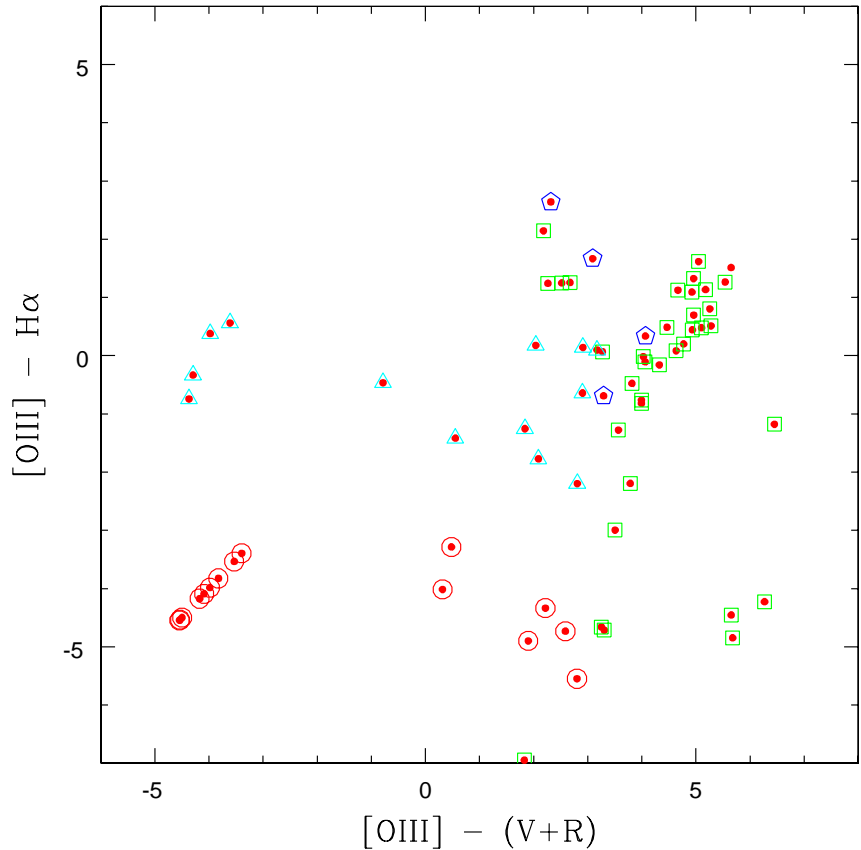}
\caption{Two-color diagram for the M86 matched sample. 
Red dots are [OIII] sources matched to the JCF90 sample. Cyan triangles
(13 objects) indicate the best PNe candidates, red circles (15) indicate the 
single line emitters with $m(V+R) > -2.8$, most likely
high excitation PNe in this situation.
Green square (32) indicates
objects with significant flux in the (V+R) image, i.e. 
$m(V+R) < -2.8$. Of the objects with blue pentagons (4), those
at $([OIII] - H\alpha) > 1$ are candidate compact HII regions.
Those at $([OIII] - H\alpha) < 1$ (2) have unreliable colors. See
discussion in the text.
\label{fig4}}
\end{figure}

For M86, the galaxy falls partially out of the frame so some candidates 
from JCF90 are outside our frame (roughly 1/3 of the total sample).
For some of the M86 objects, our photometry may not be as accurate as
for the M84 field, because of
filter edge effects and partial mosaic coverage.
JCF90 reported 140 candidates in total, and 69 were matched with Subaru
sources. The faintest PN in the matched sample has 
m(5007) = 27.94\footnote{There are only 9 PNe candidates in the JCF90
list whose m(5007) is fainter than 27.94.}, from JCF90 photometry.
5 objects of the matched list fall in bad areas of either the
$H\alpha$ or the (V+R) image, and their colors are then spurious.
Of the remaining 64 objects, 
\begin{itemize}
\item 13 are detected in [OIII] and have a positive H$\alpha$ flux,
with $m(V+R)>-2.8$ (most likely true PNe);
\item 15 are detected only in [OIII], with a negative measured
H$\alpha$ flux, and have $m(V+R)>-2.8$ (probable high excitation PNe);
\item 32 objects have $m(V+R)<-2.8$. 
21 of these fall in the area of the two-color plot where flat-spectrum 
continuum objects are found;
2 are single-line emitters, according to criterion n. 2, 
with no significant positive flux in H$\alpha$;
4 objects are most likely continuum objects with steep blue 
continuum, because their measured $([OIII]- H\alpha) < 0.0$.
We also have 5 candidates for compact HII regions: these 
objects fall in the same region of the two-color diagram as the 
possible HII region in M84, with
$([OIII]- H\alpha) > 1.0, ([OIII]-(V+R)) \simeq 3\,\, \mbox{and}\,\, m(V+R)
<-2.8$; 
\item 2 possible compact HII regions with $m(V+R) > -2.8$.
\item The remaining two objects are scattered in the region where 
the continuum objects are found, but their $m(V+R)>-2.8$.
\end{itemize}
See Figure~\ref{fig4} for the two-color plot for the M86 matched sources. 

{\it Calibration of the Subaru [OIII] fluxes --}
From the selected most likely [OIII] line emitters in M84 and M86, 
we can calibrate the Subaru [OIII] fluxes. 
We derive the calibration $$m(5007) = m([OIII]) + 26.30\pm0.05 $$ where 
$m([OIII])$ is the extinction corrected instrumental magnitude, and
$m(5007)$ is the standard [OIII] magnitude. The standard deviation
of the difference $m(5007) - m([OIII])$ for the matched sample is 
$\sigma = 0.45$ on a sample of 74 objects.  
The [OIII] magnitude $m([OIII]) = 0.7$  corresponds to $m(5007) =
27.0$, which is similar to the magnitude targeted so far for
spectroscopic follow-up.  

We checked the consistency of our [OIII] calibration by computing
the narrow band fluxes in the AB magnitude system, and tested them against the 
calibration of the V band image. 
We computed the conversion from $m(5007)$ to $m_{AB}$ magnitude: for
$\lambda_c = 5021$ \AA, $\Delta\lambda = 74$ \AA, $m_{AB} = m(5007) - 2.49$.
Therefore $$m_{AB} = m([OIII]) + const_{AB} = m([OIII]) + 23.81.$$
Following the same strategy as in the LALA survey (Rhoads et
al. 2000), we use
the V-band zero point from \S2 and set $([OIII] - V) = 4$ for blue stars.
The inferred value for $const_{AB}$ is then 24.00, with a $0.19$
difference from the value obtained via the comparison with the JCF90
sample.  Given the relative imprecision of this estimate, the agreement
with the estimate via the M84/M86 PNe is acceptable.

The results from the matched Subaru-JCF90 sample
indicated that about 38\% of the JCF90 candidates, 
which were identified via blinking and visual identifications,
have a positive detected continuum flux in our combined (V+R) image 
(25\%), or are not present in the deep Subaru [OIII] image (13\%).
This fraction (38\%) of likely spurious candidates is consistent with
the confirmed fraction of true line emitters (about 60\%)
found in spectroscopic follow-up of similarly selected intra-cluster
candidates at the 2dF (Freeman et al. 2000 ), and also in the multi-object
spectroscopy of PNe in individual galaxies,
e.g. NGC~1399 and NGC~1316 (Arnaboldi et al. 1994, Arnaboldi et al. 1998).
However, we note that problems in positioning fibers and slit masks
could also reduce the detection rate of candidate objects in the
spectroscopic programs.

\subsection{ICPN Candidates}
From the two-color plot of the whole Subaru field, we constructed a catalog
of {\it likely} PNe from objects with [OIII] detections, 
measured H$\alpha$ flux, and $m(V+R) > -2.8$. 
In the area of M84, additional 8 likely PNe were found.
We then masked regions centered on M86, M84 and NGC~4388 
with areas of $4 R_e \times 4 R_e $, to avoid the PNe candidates 
belonging to these galaxies. 
Down to our completeness magnitude limit of $m(5007)= 27.3$, 
we selected the most likely ICPNe candidates according to criterion
n. 4, and then inspected their images visually: this gave a sample of 
32 ICPN candidates. 

\begin{figure}
\epsscale{0.8}
\plotone{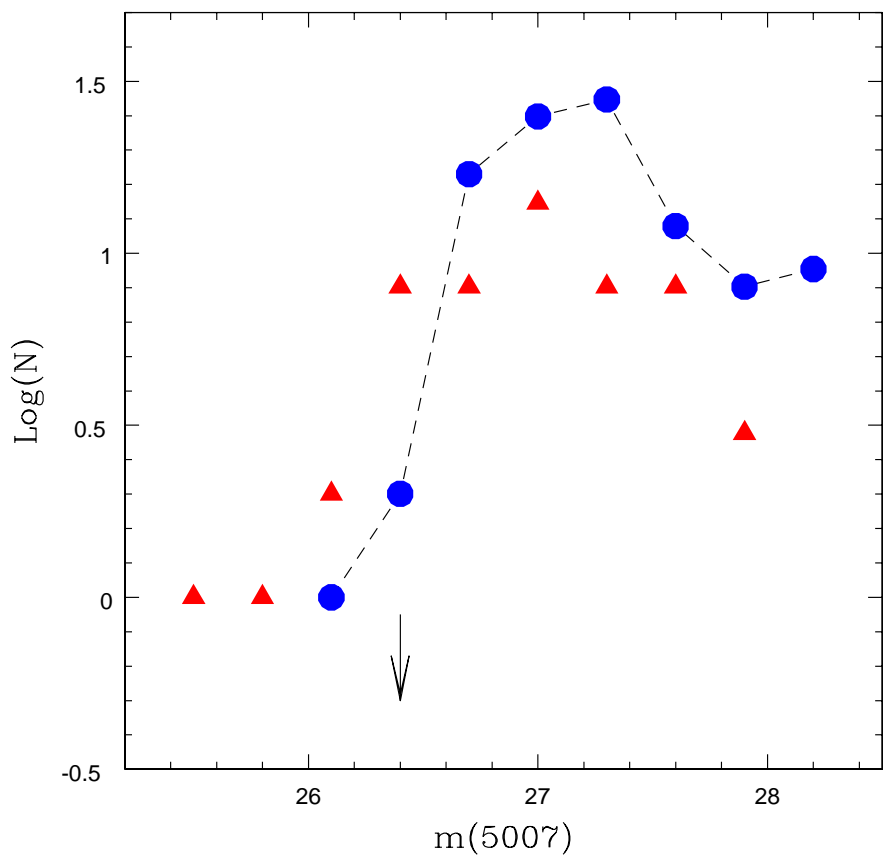}
\caption{The raw PNLF for the whole M84 PNe sample from the Subaru survey 
(red triangles) compared with the PNLF from JCF90 sample of 102 
objects (blue dots), binned into intervals of 0.3 mag.
Object n. 103 at m(5007)=24.6 is not included in the plot. The arrow 
indicates the PNLF cutoff at $m(5007) \simeq 26.4$ 
as determined from JCF90, based on the total PNe sample obtained from
their survey of Virgo giant early-type galaxies. It shows the
presence of objects that are distinctly brighter than the cutoff,
as already noted by JCF90.
\label{PNLFM84}}
\end{figure}

Based on our photometry and $m(5007)$ flux calibration\footnote{Our
[OIII] photometry is slightly biased from the contribution of the
[OIII] emission at 4958.9 \AA\, which enters in our filter bandpass 
for radial velocities $v_{rad}> 1512$ km s$^{-1}$. This may cause a 
maximal brightening of the PNLF of $\sim 0.13$ mag, see discussion
in Arnaboldi et al. (2002).}, 
in the M84 whole PN sample of 53 objects (45 matched with JCF90 + 8
Subaru PNe), there are 6 PNe (5 from JCF90 + 1 from Subaru PNe) 
which have
$m(5007) < 26.4$, see Figure~\ref{PNLFM84} where we plot the M84 PNLF
from ours and the JCF90 sample.
PNe which are found in Virgo galaxies and are brighter than 
the cutoff magnitude $m(5007) = 26.4$ were 
considered over-luminous objects by  JCF90, and,
they must be ICPN candidates on the near side of the
cluster, according to Ciardullo et al. (1998).\\
Including the over luminous objects in the Virgo galaxies, 
we then have a sample of 38 ICPN candidates, associated with the diffuse
light in the Subaru field. The complete list of candidates, their
coordinates, $m(5007)$ magnitudes, and their relative position
on the [OIII] narrow band mosaic image are available in 
Okamura et al. (2002). 

We also inspected the IC field for possible compact HII regions, 
which we can identify from our two-color diagram. 
In the IC region, we identified 3 candidate compact HII regions, 
one with $m(V+R) < -2.8$ and 2 with $m(V+R) > -2.8$. 

\section{Spectroscopic Observations}
\subsection{The M84 field}
The spectroscopic observing run in the M84 field, 
centred at $\alpha(J2000)\, 12:24:57,\, \delta(J2000)\, 12:52:25$,
was carried out at the 4 meter Telescopio Nazionale Galileo (TNG) 
on the nights of the 13-15 of April 2002, with the DOLORES
spectrograph, in multi-object spectroscopy.
The DOLORES spectrograph is equipped with a Loral thinned $2048\times 2048$
CCD, with a scale of $0''.275$ pix$^{-1}$, which yields a total
field of view of $9.4\times 9.4$ arcmin. Spectroscopic observations were
carried out with the medium resolution blue grism, 
with a dispersion of $1.7$\AA\ pix$^{-1}$ and a wavelength 
coverage $3500 - 7000$ \AA.
A mask was produced in blind offset mode, from the selected
object catalog, based on the mosaic astrometry. 
Field stars from the Subaru narrow band [OIII] image were used 
for mask positioning and alignment.  
Slitlets were positioned on the objects listed in Table \ref{tab1}.
These objects are: \\
i) all brighter than $m(5007) = 27.0$, which is roughly the
faintest magnitude at which [OIII] line emission from a PN at
the distance of Virgo can be detected spectroscopically with a 4 meter 
telescope (Freeman et al. 2000),\\
ii) have  $([OIII] - (V+R)) < 3.2$, $m(V+R) > -2.8$ and 
$-3 < ([OIII] - H\alpha) < 1$.\\
Unfortunately, weather conditions were poor during the run, with a
seeing of $1''.5-2''.0$. The total 
exposure time for this mask configuration was $10 \times 2000s$.\\
Data reduction was carried out using standard tasks in IRAF, for 
standard CCD data reduction (bias subtraction and flat fielding),
and extraction, wavelength calibration, plus sky subtraction
of two-dimensional longslit spectra. Because of bad weather
condition, flux calibrations were not acquired. 

\begin{figure}
\epsscale{1.0}
\plottwo{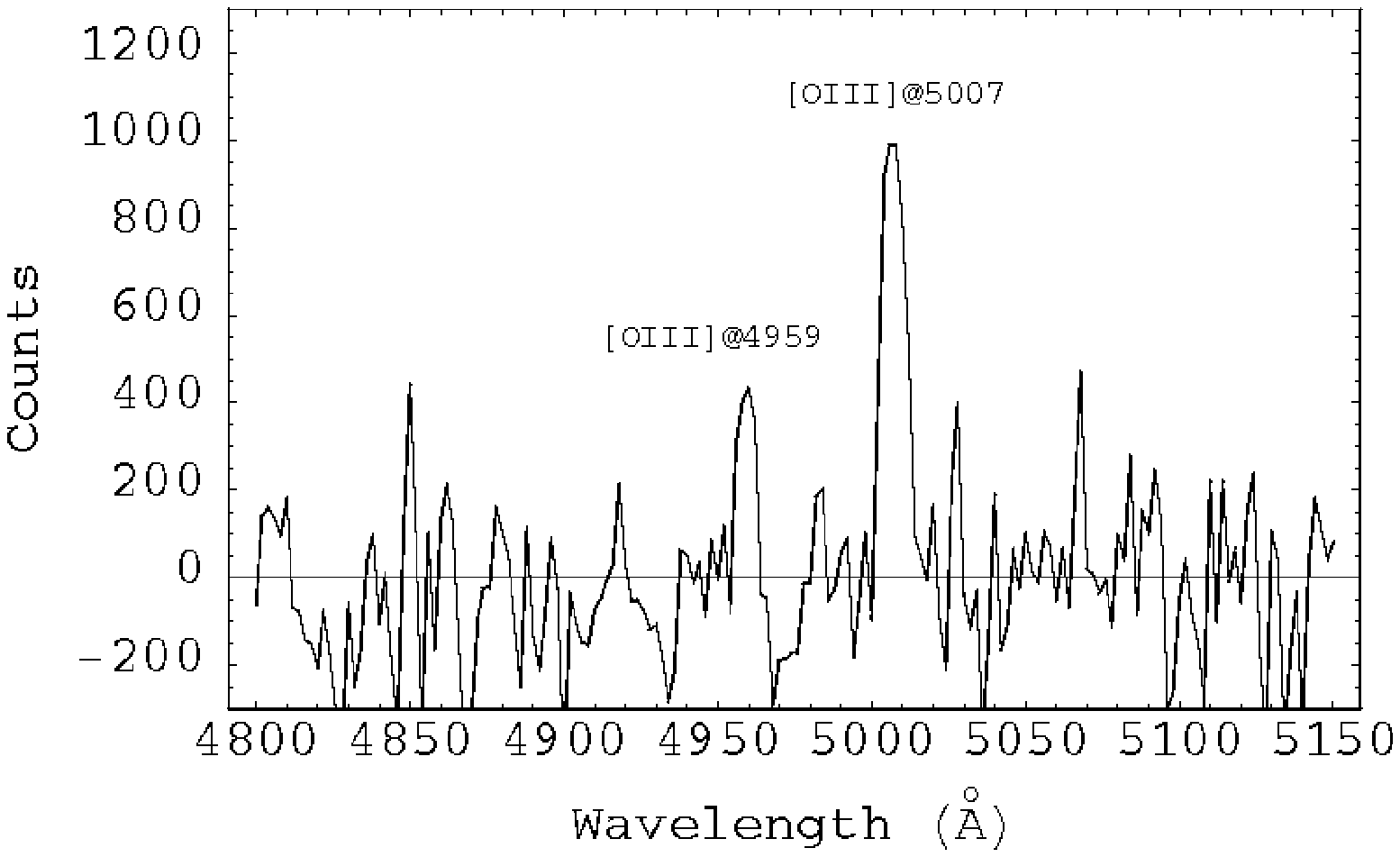}{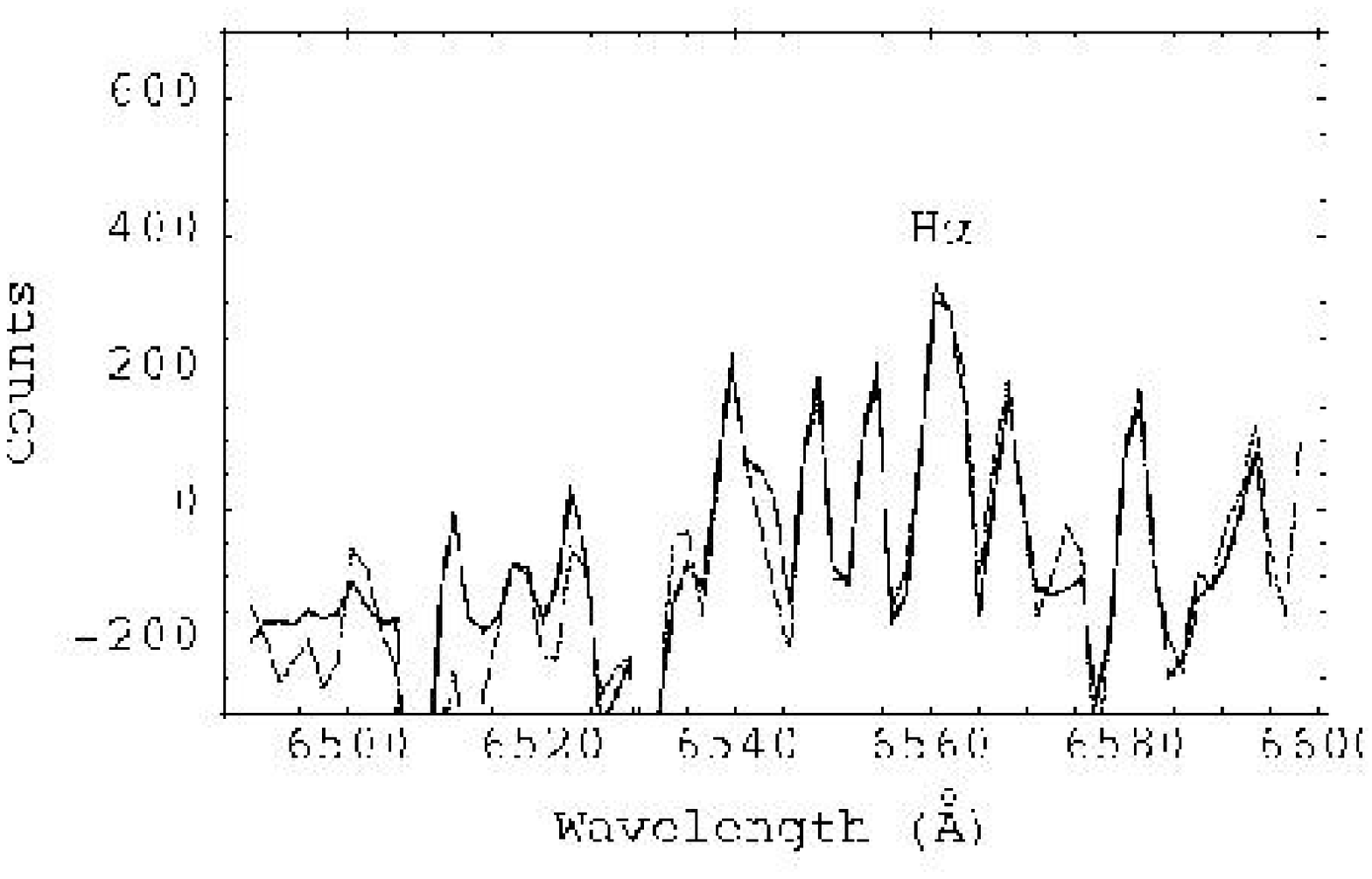}
\caption{a): spectrum obtained as sum of the 7 sharp line emission
spectra from the TNG spectroscopy, at the wavelength of the 5007 \AA\ emission.
b): in red we plot the spectrum obtained as sum of the 7 sharp line emission
spectra from the TNG spectroscopy, centred on the H$\alpha$ emission. 
In Figure 6b,  the spectrum obtained as sum of all the 8
spectra with emission around 5023 \AA\ is shown with a black continuum
line, and the summed spectrum from the 7 sharp line emitters is
plotted with a red dotted line: there is no improvement in the S/N of 
the H$\alpha$ line, therefore the asymmetric line emitter
is most likely a Ly$\alpha$. The [OIII] doublet is clearly evident 
on Figure 6a. The H$\alpha$ emission is visible in Figure 6b at similar S/N 
as the [OIII] $\lambda 4959$ \AA\ emission.
\label{TNGspec}}
\end{figure}

{\it TNG spectroscopic confirmations --}
Of the 10 objects which were allocated a slit, objects n. 103 and 
n.105 were not detected: given their bright [OIII] flux 
we believe that non-detection may be due to astrometric problems
and/or incorrect positioning of the objects in the slits.
The other 8 objects are all probable PNe with narrow line
emission in the wavelength range of the [OIII] filter, except for 1016
which shows an asymmetric broad line.
To confirm the identification as PNe, we shifted each of the 7 
likely PNe spectra so that the emission line is at the 5007\AA\ rest 
wavelength, normalized each spectrum to the same total flux in the
line, and then summed the spectra. The summed spectrum is shown in 
Figure~\ref{TNGspec}, in which both the 4959\AA\ and 5007\AA\ [OIII] 
lines are clearly visible.
Furthermore, the H$\alpha$ emission is
also visible in the red part of the summed spectrum of the
sharp line emitters (therefore true PNe), although with a worse S/N ratio 
than the [OIII] emission at 4959 \AA.
This confirms the nature of these objects as PNe.
Their mean velocity (1174 km s$^{-1}$) and velocity dispersion (317 km
s$^{-1}$) are consistent with most or all being members of M84.

The spectroscopic observations of the PNe in the outer region of M84
confirms the validity of using both H$\alpha$ and [OIII] imaging to
detect PNe near the bright cutoff of the PNLF.  The analysis of the
dynamics of the outer regions of M84 will be discussed elsewhere.

\subsection{Virgo IC field}
The spectroscopic follow-up of ICPN candidates in the intracluster
field centred at $\alpha(J200) 12:25:31.9,\, \delta(J2000) 12:43:47.7$
was carried out as a backup program at UT4 of the VLT at Paranal, 
on the nights of the 13-14 of April 2002, with FORS2, in MOS mode.
The FORS2 field of view covers an area of $6.8\times 6.8$ arcmin
in standard resolution. 

Our selected targets were each assigned a MOS slitlet
(in MOS mode, up to 19 movable slitlets can be
allocated within the field). Observations were carried out with
GRISM-150I and the order separation filter GG435+81, giving a wavelength
coverage of $4500 - 10200$ \AA\ and a dispersion of $6.7$ \AA\ pix$^{-1}$.
The angular scale across the dispersion is $0.126$ arcsec pix$^{-1}$. The
nights were clear but not photometric.
The mean seeing was better than 1.0 arcsec despite strong northerly
winds.
Because the observations were done in blind offset mode,
the slitlet width was chosen to be $1.4$ arcsec. Slitlets were positioned
on the three objects listed in Table \ref{tab2}.
Three stars in the field were selected for pointing checks, 
and two stars were used to obtain the correct 
mask position and alignment, via the acquisition of their
``through slit'' image. 
The total exposure time was $7\times1800s$. 
Spectrophotometric standard stars were 
observed at the beginning and end of the nights, but
the conditions made the flux calibration uncertain.

{\it Selected objects in the IC field --}
According to the selection criteria discussed in
Section \ref{color}, IC1 is a likely compact
HII region: it is bright in [OIII] ($m([OIII]) = -0.62$),
H$\alpha$ ($m(H\alpha)=-1.91$), and also in (V+R) $(m(V+R) = -3.95)$. 
IC2 and IC3 are ICPN candidates; they are brighter than 
$m(5007) = 27.0$, and have $([OIII] - (V+R)) < 3.2$, $m(V+R) > -2.8$
and $-3 < ([OIII] - H\alpha) < 1$.
Note that from Figure~\ref{fig2} we may still have some contaminant 
Ly$\alpha$ emitters in our best sample, because both 1) our off-band
image and 2) the H$\alpha$ image are not deep enough. 
The H$\alpha$ image of the field is shown in Figure~\ref{ICfield}.

{\it VLT spectroscopic confirmations --}
We were able for the first time to acquire a spectrum of a single ICPN
with enough S/N to show both of the [OIII] lines clearly.  The 
spectrum of this object (IC2) is shown in Figure~\ref{ICPNsub}a.  The
observed flux ratio of the two [OIII] lines is 3.1, and the
[OIII]/(H$\alpha$ + [NII]) is 4.6, as expected for a high excitation
PN. \\
Object IC1 is indeed a compact HII region: its spectrum, in the 
wavelength region $ 4000 - 9500$ \AA, and the physical properties 
will be discussed in detail in Gerhard et al. (2002).\\
IC3 turned out to be a single line emitter, probably
a Ly$\alpha$ object at z=3.13, see Figure~\ref{ICPNsub}b.
A very weak continuum is visible in the VLT deep spectrum,
so this object appears similar to some of the blue
continuum contaminants discussed earlier.\\

\section{Fraction of diffuse light in Virgo cluster core}
Our goal is to estimate the fraction of light from intracluster
stars in the Virgo cluster core. We first
estimate the total intracluster stellar luminosity associated with our
ICPNe by computing the number of planetary nebulae per unit luminosity
$\alpha_{1.0}$ within 1.0 mag of the bright cut off of the [OIII] PNLF,
using the PNe and background light of M84 with our 
sample\footnote{This approach is not ideal, because the 
estimated $\alpha_{1.0}$ values depend on the integrated B-V color of
the underlying population: see Hui et al. (1993) and reference therein.}  

The M84 data are particularly relevant in our discussion as the galaxy is 
well inside our Subaru field.  We can use its PN population to
evaluate the specific frequency parameter 
$\alpha_{1.0}$ (i.e. the number of PNe per unit 
luminosity in the first 1.0 mag of the [OIII] PN luminosity function),
as selected according to our criteria.  

In the region of M84 with $R > 1 R_e$,
we detected 45 PNe in M84 ( 40 from JCF90 + 5 TNG confirmed PNe from
Subaru sample) within $\sim 1.0$ mag of the PNLF cutoff at 
$m(5007) \simeq 26.4$. 
Adopting  $M_{B_T} = -20.79$ for this galaxy, the total blue light from 
the region beyond $1R_e = 57$ arcsec is 
$1.3\times 10^{10} L_{B,\odot}$, which leads to
$$\alpha_{1.0} = 3.46 \times 10^{-9}~{\rm PN~L}_{B,\odot}^{-1}.$$

The 36 ICPN candidates\footnote{IC3 and Object n. 103, both
in our ICPN sample, were not confirmed as such in the spectroscopic follow-up}
in our Subaru field imply a total associated
luminosity of $1.0 \times 10^{10}L_{B,\odot}$.  The area surveyed by
the Subaru field is 0.196 deg$^2$ and the associated luminous density
of the intracluster medium is then $1.4 \times 10^7$ L$_{B,\odot}$
arcmin$^{-2}$ or $0.76 L_{B,\odot}$ pc$^{-2}$. This corresponds to a
surface brightness of $\mu_B = 27.4 $ mag arcsec$^{-2}$.  Some fraction
of our 36 ICPN candidates may still turn out to be high redshift
emitters. For this relatively bright sample, selected from [OIII] and
H$\alpha$ observations, the fraction of high redshift objects is
probably less than the 25\% found by Arnaboldi et al. (2002),
based on Freeman et al. (2000) spectroscopic follow-up,
for an [OIII]-selected sample. 
In order to derive a lower limit on the
contribution of the diffuse light, we will assume that 75\% of our 36
ICPN candidates are indeed ICPNe, reducing the inferred surface
brightness of the diffuse light in our field to $\mu_B = 27.7 $ mag
arcsec$^{-2}$, and the total associated luminosity to $7.5 \times
10^9L_{B,\odot}$.

We now need to compare this with the luminous contribution from the
Virgo galaxies.  What is the most appropriate estimate to compare with
the diffuse stellar density ?  We distinguish two extremes:\\
A) if the IC stellar population is related to the local density of
galaxies, then we should compare its luminous contribution to that of
the galaxies in the field. The three galaxies M86, M84 and NGC 4388
dominate the galaxy light in our field, and contribute a total light of
$7.2 \times 10^{10} L_{B,\odot}$ (total magnitudes from NED).
Therefore in this field the percentage of diffuse light is at least
10\% of the light from the cluster galaxies.\\
B) If the IC stellar population is a larger-scale phenomenon, as it
would be if it arises from galaxy harassment, then we should compare
its surface brightness with the smoothed-out surface brightness of
galaxies (derived from the large-scale distribution of galaxy light in
the cluster) at the radius of our field.  Binggeli et al. (1987) fit a
King model to the luminosity-weighted number counts of galaxies in the
Virgo cluster, and derive a (distance independent) central surface
brightness of $6.9 \times 10^{11} L_\odot$ Mpc$^{-2}$ and a core radius
$r_c = 1.7$ degrees for the cluster. Our field is within $0.25 r_c$ of
their adopted center. The smoothed-out surface brightness of galaxies
in our field is then $\mu_B = 27.4 $ mag arcsec$^{-2}$ and the diffuse
light (with $\mu_B = 27.7 $ mag arcsec$^{-2}$) contributes about 40\%
of the total stellar light (galaxies + diffuse).\\
For a detailed discussion of the lower limit on the baryonic fraction 
in the Virgo cluster core see Okamura et al. (2002).

\section{Conclusion}
We have used H$\alpha$ and [OIII] narrow band images from the Suprime-Cam
prime focus camera of the 8.2 meter Subaru
telescope to distinguish bona-fide planetary nebulae from high 
redshift emission line galaxies at the bright end of the [OIII] 
luminosity function. 
The selection criteria were derived from the two-color diagram, 
([OIII] -- H$\alpha$) vs. ([OIII] -- (V+R)): simulated populations of 
PNe, single-lined emitters and continuum sources are constructed 
in the two-color diagram, and are
used to outline the regions inhabited by the different kinds of
objects.  These criteria were then calibrated against 
{\it i)} previously discovered PNe in the 
galaxies M84 and M86 by Jacoby et al. (1990), and {\it ii)} 
subsequent spectroscopic follow up. 

The spectroscopic follow-up of a subsample in the M84 field
was carried out at the TNG, with the DOLORES spectrograph.
8 objects were confirmed as line-emitters; the summed spectrum of
the 7 sharp-line emitters shows the $\lambda 4959$ [OIII] 
and the H$\alpha$ emission, in addition to the PN strongest $\lambda
5007$ [OIII] emission. The broad line emitter is most likely a 
Ly$\alpha$ at redshift 3.13.

The spectroscopic follow-up of a subsample in the IC field
was carried at UT4 of the VLT, with FORS2.
These observations gave us the first unambiguous confirmation of 
the presence of an ICPN.
A high S/N spectrum for candidate IC2 was obtained at the VLT which
clearly shows the presence of both the [OIII] $4959$ and
$5007$\AA\ lines, and the H$\alpha$ emission.

Based on these criteria, we have identified a sample of 36 ICPNe 
candidates in the effective survey area of $0.196$ deg$^2$, in the Virgo core. 
This would correspond to a surface brightness
of $\mu_B = 27.4 $ mag arcsec$^{-2}$ for the underlying diffuse
intracluster stellar population. With some allowance for residual
contamination of our sample by high redshift emitters, we derive lower
limits to the fractional contribution of the diffuse intracluster light
between 10\% and 40\% of the total light (galaxies plus diffuse light),
depending on the assumptions about the nature of the intracluster
stellar population.  If it is associated with the harassment process,
then the 40\% limit is more appropriate.  These estimates neglect
substructure of the diffuse medium on scales larger than about $0.5$
degrees. We cannot refine these limits until more is known about the structure
of the diffuse intracluster stellar medium. This is the longterm goal
of our program.

\acknowledgements M. A. and K.C.F. wish to thank the NAOJ staff for
their help and support during their visit to NAOJ in November 2001, and
acknowledge the support from S.O. research funds for travel and living
grants while in Japan.  M.A. and O.G. wish to thank 
R. Scarpa, N. Patat for their help and support during the mask
preparation in blind off-set mode with FORS2 in MOS mode, and 
the observations at UT4. M.A., N.N., and M.P. wish to
thank E. Oliva, E. Held, G. Covone and the TNG staff for the help and support
during mask preparation and the observations carried out at TNG.
The authors wish to thank T.Hayashino, H.Tamura, and Y.Matsuda for
their help in measuring the characteristics of the narrow band
filters, and M.A. Dopita for useful discussions on PNe 
luminosity function. 
S.O. acknowledges the Grant-in-Aid (13640231) from the Ministry of 
Education, Culture, Sports, Science and Technology in support of this
research.
O.G., K.C.F, and M.A. wish to thank the Swiss National Foundation for 
research money and travel grants.
Data reduction was carried out at the computer system operated by
Astronomical Data Analysis Center of the National Astronomical
Observatory of Japan. 
This research has made use of the NASA/IPAC Extragalactic Database
(NED) which is operated by the Jet Propulsion Laboratory, California 
Institute of Technology, under contract with the National Aeronautics
and Space Administration.

\appendix

\section{Converging beam correction to the filter transmission curve}

{
The converging beam from any telescopes will alter the transmission
properties of an interference filter. If the f-ratio of the telescope
is slower than about f/8, the effects can generally be
ignored. However the Subaru/Suprime camera is in a very fast f-ratio (1.86)
and the filter transmission curve can be affected by it.
We provide the filter characteristics which were measured by T.Hayashino,
H.Tamura, and Y.Matsuda in a converging beam with the same f-ratio as the
Subaru prime focus at 13 different positions almost uniformly located
over the effective surface of  the filter. The beam size on the
filter surface, 30mm in diameter, is also the same as that at the
Subaru prime focus (Hayashino et al. 2003).
Figure \ref{filtervirgo} shows the filter transmission curves at the
13 positions. One curve shifted toward short wavelengths is for the
very central position.
}

\begin{figure}
\epsscale{0.8}
\caption{H$\alpha$ image of a $ 4 \times 3$ arcmin field in the Virgo IC
region. This is part of of the region selected for the spectroscopic 
follow-up with FORS2 at UT4 of the VLT; North is up and East to the left.
The confirmed ICPN (IC2) is circled in yellow and red, in the upper
part of the image. The confirmed compact HII region (IC1) is the
object circled in blue, on the left portion of the image. 
The two objects circled in red (in the central region of the image and
in the lower right corner of the image) are two additional ICPN candidates,
which will be observed in a subsequent spectroscopic follow-up. 
\label{ICfield}}
\end{figure}

\begin{figure}
\epsscale{1.3}
\plottwo{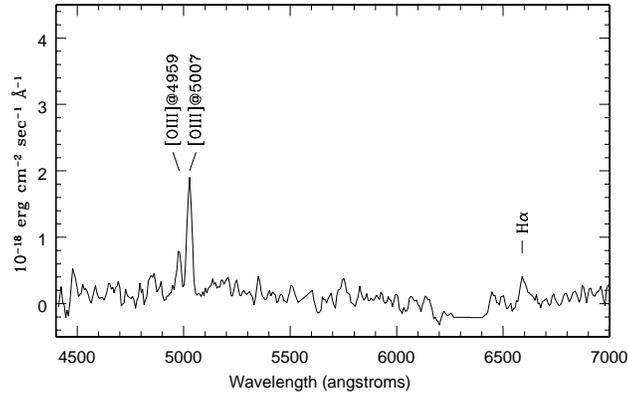}{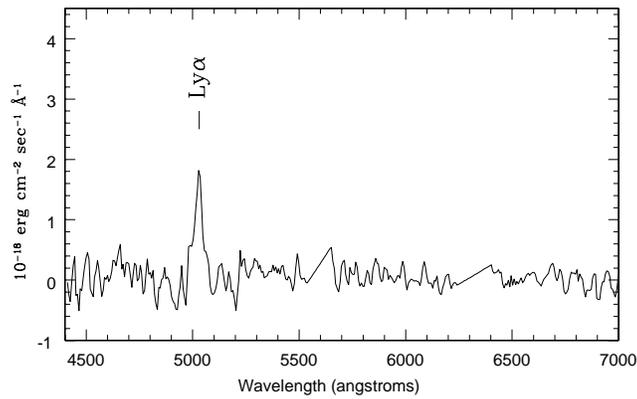}
\caption{a): 1-D spectrum of IC2, a confirmed intracluster PN in
the Virgo cluster. The [OIII] doublet and the H$\alpha$ emission are 
visible with a large S/N. b): 1-D spectrum of IC3, a single 
line emitter, with a broad-asymmetric line. This is most likely 
a Ly$\alpha$ emitter at redshift 3.13.
\label{ICPNsub}}
\end{figure}

\begin{figure}
\epsscale{0.8}
\plotone{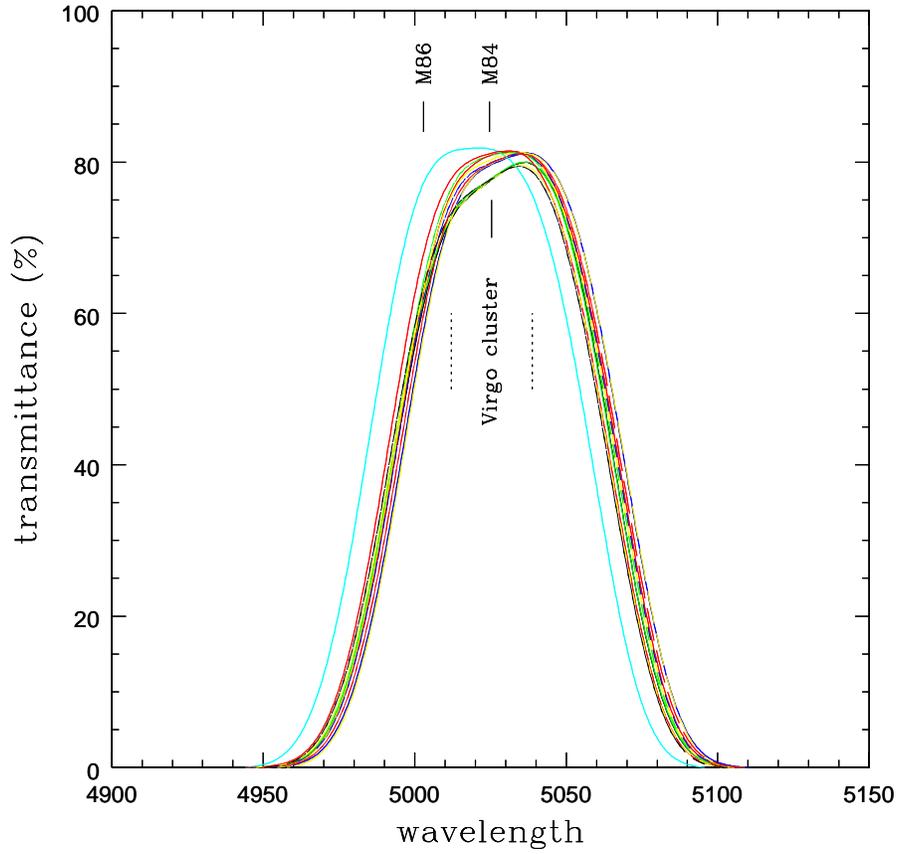}
\caption{The [OIII] filter transmission curve measured in a f/1.86
converging beam at 13 different positions. The systemic velocity of
M84, M86 and of the Virgo cluster are marked. The dotted marks
indicate the $\pm 800$ km s$^{-1}$ of the velocity dispersion of Virgo
cluster galaxies
\label{filtervirgo}}
\end{figure}

\clearpage

\begin{table}
\caption{\label{log} Observing log of the Subaru imaging run}
\begin{tabular}{crrrc}
\tableline
Band&\multicolumn{2}{c}{Exposure (sec)} &seeing&Date \\
\tableline
\tableline
$V$ & 900 &(300$\times$3)& 0.85-1.00 & 24 Mar., 2001\\
$R$ & 720 &(240$\times$3)& 0.75-0.98 & 24 Mar., 2001\\
{[OIII]} & 3600&(1200$\times$3)& 0.65-0.68 & 23 Apr., 2001\\
H$\alpha$& 8728& (1200+328)& 0.62-0.70 & 24 Apr., 2001\\
         &     &(1200$\times$6) &  0.62-0.68 & 25 Apr., 2001\\
\tableline
\end{tabular}
\end{table}

\clearpage

\begin{table}
\caption{\label{tab1} Objects selected according to their 
[OIII] and H$\alpha$ flux in the outer region of M84. 
$\lambda_{obs}$ is the wavelength of the main emission 
at the green wavelengths, v$_{rad}$ is the radial velocity determined
for the sharp line emitters. See discussion in the text.
}
\begin{tabular}{cccccc}
\tableline
IDENT  &    RA(J2000)	& DEC(J2000)	& $m([OIII])$ & $\lambda_{obs}$ & v$_{rad}$\\
\tableline
\tableline
103\tablenotemark{1}  & 12:24:53.835 & 12:52:25.12 &  -1.6967  & & \\
105  & 12:25:04.765 & 12:54:29.81 &   0.1287  & & \\
106  & 12:24:52.962 & 12:49:25.40 &   0.1441  & 5018.2 & 671.061 \\
107  & 12:25:03.916 & 12:55:25.50 &   0.2314  & 5023.6 & 991.612\\
109  & 12:25:08.259 & 12:54:37.99 &   0.2318  & 5025.3 & 1096.46\\
1010 & 12:25:07.450 & 12:51:02.98 &   0.3990  & 5034.2 & 1629.72\\
1011 & 12:24:55.613 & 12:53:10.00 &   0.4181  & 5031.0 & 1438.0\\
1013\tablenotemark{2} & 12:24:54.59  & 12:52:39.89 &   0.4601  & 5029.1 & 1324.151\\
1016 & 12:25:01.915 & 12:49:48.22 &   0.5054  & 5026.8 & \\
1017\tablenotemark{2} & 12:25:12.776 & 12:53:34.69 &   0.5787  & 5024.8 & 1066.51\\
\tableline
\end{tabular}
\tablenotetext{1}{Object n.103 has m(5007) = -1.7 + 26.3 = 24.6, 
so it is a ICPN candidate} 
\tablenotetext{2}{Objects n.1013 and n.1017 are in common with JCF90} 
\end{table}

\clearpage

\begin{table}
\caption{\label{tab2} Objects selected according to their 
[OIII] and H$\alpha$ flux in the IC field at 
$\alpha(J200) 12:25:31.9,\, \delta(J2000) 12:43:47.7$. 
$\lambda_{obs}$ is the wavelength of the main emission 
at the green wavelengths. See discussion in the text.
}
\begin{tabular}{ccccc}
\tableline
IDENT  &    RA	&	DEC	& $m([OIII])$ & $\lambda_{obs}$\\
\tableline
\tableline
IC1  & 12:25:43.11 & 12:43:06.58  &  -0.62  & 5051.0 \\
IC2  & 12:25:35.86 & 12:44:52.09  &   0.34  & 5027.6 \\
IC3  & 12:25:20.61 & 12:44:45.44  &   0.41  & 5029.1 \\
\tableline
\end{tabular}
\end{table}

\end{document}